# Tracking the performance of an R&D programme in the Biomedical Sciences

**Nicolas Robinson-Garcia[1,2]\*, Alvaro Cabezas-Clavijo[1] and Evaristo Jiménez-Contreras[1,3]**

*1 EC3metrics SL spin-off, Universidad de Granada, Spain, 2 INGENIO (CSIC-UPV), Universitat Politècnica de València, Camino de Vera s/n, 46022 Valencia, Spain, 3 Departamento de Información y Comunicación, Grupo de investigación EC3, Universidad de Granada, Spain*

**\* Corresponding author:** elrobin@ingenio.upv.es

This paper aims at offering an evaluation framework of an R&D programme in the Biomedical Sciences. It showcases the Spanish Biomedical Research Networking Centres initiative (CIBER) as an example of the effect of research policy management on performance. For this it focuses on three specific aspects: its role on the national research output in the biomedical sciences, its effect on promoting translational research through internal collaboration between research groups, and the perception of researchers on the programme as defined by their inclusion of their CIBER centres in the address field. Research output derived from this programme represents around 25% of the country's publications in the biomedical fields. After analysing a seven year period, the programme has enhanced collaborations between its members, but they do not seem to be sufficiently strong. With regard to the credit given to the initiative, 54.5% of the publications mentioned this programme in their address, however an increase on the share of papers mention it is observed two years after it was launched. We suggest that by finding the point in which the share of mentions stabilises may be a good strategy to identify the complete fulfilment of these types of R&D policies.

**Funding**

This work was supported by the Spanish Fondo de Investigación Sanitaria (FIS) [PI10/01122].

## 1. Introduction

Efforts on linking basic research with clinical practices in the biomedical sciences has lately become a priority issue in the research agenda of many countries (Barjak, Es-Sadki, & Arundel, 2015; Rettig, Schechter, & Perlman, 2004). Since 1979, many have denounced a declining interest on clinical research and on the translation of basic research to societal demands (i.e., Rettig et al., 2004). As a consequence, in the last decade some countries have introduced research policies to enhance 'translational' research and promote collaboration between clinicians and researchers. Examples of such policies are the European Clinical Research Infrastructures Network (Demotes-Mainard & Ohmann, 2005) or the consortium of Clinical and Translational Science Centers promoted by the NIH in the United States (Butler, 2008). The basis of such programmes rests on the idea that by strengthening collaboration between researchers and practitioners the interactions between them will eventually lead to translational outcomes.

While this approach seems reasonable, it will succeed depending on the role collaboration plays as a catalyst between research output and clinical practice. In this sense, many studies have explored publication patterns related with the production of translational research in the biomedical sciences. Luwel and van Wijk (2014) analysed biomedical journals indexed the Web of Science database containing the word 'translational' in their title and comparing them with other biomedical journals. Their goal was to study if they published more interdisciplinary work than other journals. They concluded by observing that interdisciplinary work took place more often in papers co-authored by researchers from different institutional categories. Lander and Atkinson-Grosjean (2011) deepened on the relationship between scientists and clinicians when researching on a particular disease. Among other findings, they reported that these interactions were more fruitful in the public sector and that the translational research process was 'iterative and untidy' (Lander & Atkinson-Grosjean, 2011). Molas-Gallart, D'Este, Llopis and Rafols (2014) explored the ways in which translational research takes place and what factors may promote collaboration between the different actors.

Scientific collaboration is a widely studied topic in the field of scientometrics (i.e., Beaver & Rosen, 1978; Katz & Martin, 1997). Generally, it is studied by using co-authorship as an indicative of such collaboration. Although co-authorship is not the only trace collaboration leaves, studies in this regard have shown that it enhances research productivity (Lee & Bozeman, 2005) and citation impact (Glänzel, 2001). Other studies suggest further benefits when such collaboration is between authors from different institutional categories, such as more innovative and creative research (Bordons, Aparicio, & Costas, 2013). In this context, it is not surprising the introduction of research policies and strategies to promote such collaboration links in the biomedical sciences. But still, evidences on the success of such programmes are difficult to retrieve.

This paper focuses on a specific research programme that aims to establish such collaborative networks between clinicians and researchers at the national level. Specifically, we analyse the case of the Spanish Biomedical Research Networking Centres (known as CIBER for their Spanish acronym). Researchers and policy makers in Spain have shown great interest in the last decade to promote collaboration in the biomedical field introducing many initiatives and strategies in its national research agenda (de Pablo & Arenas, 2008). Here we mention the introduction of the FIS/Miguel Servet Research Contract Programme which intends to incorporate basic researchers in hospitals (Rey-Rocha & Martín-Sempere, 2012). The CIBER initiative is the most important programme in Spain with an annual budget of around €42,000,000. This programme was launched in 2006 pursuing the following goals:

1. Promote excellent research in the biomedical sciences in the National Health System and the National Science and Technology System by launching and promoting stable networking structures.

2. Enhance collaboration links between different research groups through these networking structures in order to strengthen research conducted on the priority areas stated by the different Spanish National Research & Development Plans.

3. Promote translational research by integrating research groups and research members from different institutional categories and connecting clinical practice with basic research.

**Table 1.** Description of the nine CIBER centres, acronyms and launch year.

| CIBER Centre | Acronym | Researchers | Groups | Launch year |
|---|---|---|---|---|
| Bioengineering, biomaterials & nanomedicine | CIBER 1 | 647 | 47 | 2006 |
| Epidemiology & public health | CIBER 2 | 474 | 47 | 2006 |
| Obesity and nutrition | CIBER 3 | 484 | 27 | 2006 |
| Hepatic and digestive diseases | CIBER 4 | 555 | 49 | 2006 |
| Neurodegenerative diseases | CIBER 5 | 808 | 60 | 2006 |
| Respiratory disorders | CIBER 6 | 464 | 32 | 2006 |
| Rare diseases | CIBER 7 | 873 | 59 | 2006 |
| Diabetes and metabolic disorders | CIBER 8 | 376 | 29 | 2007 |
| Mental health | CIBER 9 | 354 | 26 | 2007 |

As a result, seven CIBER centres were created in 2006 and two more in the subsequent year. Each of the nine centres is thematically oriented and comprises a number of research groups scattered through the country. These centres are not physical, - meaning that they do not bring together geographically the research groups which are still located in their original institution, - but serve as virtual platforms by which collaboration can be channelled. In table 1 we include the fields in which each centre is focused, the year in which they were launched and the acronym by which they will be referred to in this study. Research groups integrating each centre were selected by a national open call and these could be placed in universities, public research organizations, hospitals or other research foundations (Molas-Gallart et al., 2014).

The CIBER initiative has gained great interest since its conception, not only nationally, but also at the international level, analysing different aspects of such programme. Three years after its creation, a bibliometric report focused on CIBER 2 indicated that collaboration patterns within the research groups that belonged to this centre were similar to those not belonging to the programme and the overall showed slightly lower collaboration patterns than the national average in biomedicine (Méndez-Vásquez, Suñén-Pinyol, Olivé-Vázquez, Cervelló-Gonzáles, & Camí, 2009). Contrarily, Delgado Rodríguez (2012) enhanced the importance of the CIBER programme and specifically the role that CIBER 2 played on the promotion and diffusion of excellent research in the Spanish biomedical sciences. Such optimistic view on the importance of such initiative in the field seems to be shared by most researchers and clinicians belonging to the programme (Cabezas-Clavijo, Robinson-Garcia, & Jimenez-Contreras, 2015).

Morillo, Díaz-Faes, González-Albo, and Moreno (2014) offer an interesting perspective with regard to the impact of the CIBER centres on collaboration and citation impact. Instead of focusing on specific centres, they analyse two different disciplines which should fairly represent CIBER 4 and CIBER 9, concluding that collaboration and impact rates are higher to papers produced by researchers belonging to a CIBER centre. However, in order to identify research output belonging to researchers assigned to a CIBER they use the address information. This is problematic, as they later acknowledge (Morillo, Costas, & Bordons, 2015), as researchers do not always acknowledge their affiliation to these 'virtual centres'. This aspect was later confirmed in the study conducted by Cabezas-Clavijo et al. (2015) who interviewed the directors of each centre where they acknowledged the difficulties encountered to make researchers feel part of the CIBER centres and include them in their affiliations. In order to solve such issue, Morillo et al. (2015) looked into the funding acknowledgments information provided by Web of Science and compared their capacity to retrieve CIBER outcomes with the list of disambiguated authors developed at the CWTS (Caron & van Eck, 2014) finding out that around 80% of the papers were retrieved when combining address and acknowledgments information.

This paper aims at analysing the global performance of the CIBER initiative based on its original objectives. So far, no study has done this; always focusing on one or two of the nine centres. It also intends to offer a framework that allows research policy makers to track the effect of their strategies on research outcomes as well as how such efforts are perceived by researchers through their affiliation links. Do they acknowledge the CIBER infrastructure more often in highly cited papers as suggested elsewhere (Costas & van Leeuwen, 2012) or do they do it when collaborating with other research groups from their centre? Specifically, our purpose is to answer the following research questions:

- What role has the CIBER initiative played in the Spanish research outcomes in the biomedical fields? Has it improved the productivity and citation impact of publications?

- Have these CIBER centres been able to improve collaboration links between research groups included in the programme? Can a growth in the collaboration between different institutional categories be observed through co-authorship?

- How do researchers perceive the role and influence of funding on their research activity? Do they include the CIBER affiliation only in papers published in high impact journals?

Although there are other elements rather than research output when implementing translational practices, in this paper we will tackle such an issue from a bibliometric perspective, focusing mainly on research publications and leaving aside other outcomes such as clinical guidelines, workshops, etc.

The structure of the paper is as follows. The Material and methods section describes the list of research groups and researchers received from the Instituto de Salud Carlos III which coordinates the CIBER programme. It details the data retrieval and processing as well as the identification of publications. It then describes the indicators and techniques employed to pursue the objectives of the study. The results of the various analyses are shown next, structured in three subsections each for each of the specific research questions. In the Discussion and Concluding remarks section we analyse the results obtained relating them with the objectives of the programme and the results obtained by similar studies. We also include some lessons learned on the perception researchers have of initiatives such as the one discussed here and how research policy makers can introduce policies that can better influence research outcomes alienating them with societal demands.

## 2. Material and methods

We conducted an analysis on the scientific output of researchers belonging to the nine CIBER centres during the 2005-2011 time period. We used two types of data sources: the Instituto de Salud Carlos III who provided data regarding the researchers and research groups belonging to each of the centres, and the Web of Science database in order to obtain the research publications and citation data of each researcher. In this section we first detail the retrieval and processing of the dataset employed for the analysis and we then define the indicators and techniques employed in order to undertake this study.

*2.1. Data retrieval and processing*

The Instituto de Salud Carlos III provided us with internal data regarding with the annual budget of each centre for the 2006-2011 time period, research group and centre to which it is affiliated, lead researcher of each team, institutional affiliation of the research group and Spanish region in which it is located. This dataset also included the list of researchers linked to each research group. At a first stage, they also offered information related with the research output of each researcher, but after testing such information we found out that the list of publications was incomplete.

For this reason, in January, 2012 we proceeded to download the research output of Spain during the 2005-2011 time period from the Thomson Reuters Science Citation Index database. We include the year 2005 as this would allow us to see their research performance prior to the establishment of the network infrastructure. We included only the following document types: articles, reviews, letters and editorial material. The Spanish research output retrieved for the study period is of 277,127 scientific papers. This information was later linked with journal information retrieved from Thomson Reuters Journal Citation Reports (JCR) which includes a set of bibliometric indicators useful to analyse the scientific visibility of the research outputs.

With these two datasets we proceeded to link publications to individual researchers (Figure 1). We did this by generating automatic variants of each researcher's name and crossing them with the author field. Such links were limited only for papers where the affiliation information of the two datasets coincided. Finally, we took into account the scientific areas to which these researchers belonged, deleting links to papers from scientific areas which fell apart from their line of work. This set of linked publications and researchers was manually checked by an information specialist eliminating false positives and checking for false negatives by enquiring the database with further name variants that were not considered at first.

**Figure 1.** Data processing and matching between the researchers and publications datasets

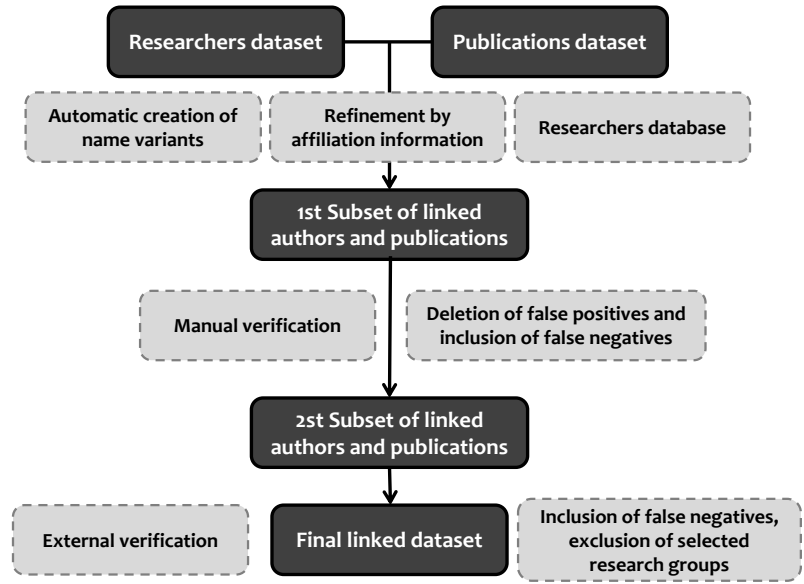

This processing work allowed us not only to link research output at the research group level, but at all levels (individual, CIBER centre and institutional), obtaining a highly grained dataset. However, in order to validate it we required external verification. This is a crucial point in our analysis as it was done at the micro and meso level, and hence errors in the dataset may invalidate the whole study (van Leeuwen, 2007). Such validation was undertaken in two stages. In a first stage, a preliminary set of results was presented to a control group of researchers affiliated to the CIBER programme who refine the set of bibliometric indicators used. Then, we reported our results to the director of each CIBER. This allowed us to exclude several research groups which were originally selected to be part of the programme but which finally were not integrated. In order to exclude their output in a systematic way, we disregarded research groups which had never collaborated with another CIBER group and had never signed their papers as CIBER during the study time period.

**Table 2.** Set of indicators used in the study

**Output**

| Indicator | Acronym | Definition |
|---|---|---|
| Number of publications | P | Publications indexed in the Science Citation Index. The considered types were articles, reviews, letters and editorial material. |

**Citation Impact**

| Indicator | Acronym | Definition |
|---|---|---|
| Number of citations | C | Total number of citation received by unit of analysis |
| Citation per paper | C/P | Ratio of citations received per paper published. |

**Visibility**

| Indicator | Acronym | Definition |
|---|---|---|
| Share of $1^{st}$ quartile papers | %Q1 | Share of papers published in journals positioned in the top 25% of their Web of Science subject category according to their Journal Impact Factor. If the journal is classified in more than one category the highest position remains. |
| Share of $1^{st}$ decile papers | %D1 | Share of papers published in journals positioned in the top 10% of their Web of Science subject category according to their Journal Impact Factor. If the journal is classified in more than one category the highest |

| | | | |
|---|---|---|---|
| **Collaboration** | | | |
| Density | | D | This indicator shows the level of cohesion between the nodes in a network. It is defined as the number of links established between nodes in relation with the highest value they could have if all nodes were connected with each other. It is a normalized indicator ranging from 0 to 1. |
| Main Component | | Co | Share of nodes connected with each other at least once in the network (meaning that they have co-authored at least one paper) in the largest cluster of the network. |
| Share of institutional class collaboration | | IC | Share of publications co-authored by CIBER research groups belonging to different institutional categories (i.e., hospitals and universities). |

*2.2. Indicators and methods*

Three sets of indicators were used in this paper: production, scientific impact and visibility, and network analysis indicators. Table 2 includes a list of the indicators as well as a definition for each of them. We calculated these indicators at different aggregation levels: Spain, all CIBER output, by CIBER centre, by institutional category and by research group.

### 3. Results

*3.1. General view of the CIBER outcome in the biomedical fields in Spain*

A total of 5010 researchers' output grouped on 376 research teams was analysed (Table 1). They produced a total of 28251 publications between 2005 and 2011. In table 3 we show their publications, citation and journal impact indicators. CIBER 2 was the most productive centre (4508 papers) followed by CIBER 1 (4411) and CIBER 4 (4356). The least productive centre was CIBER 8 (1710). Regarding their citation impact, CIBER 5 and 2 were the ones with a highest citation average (12.47 and 12.43) while CIBER 9 and 1 had the lowest citation rate per paper (9.52 and 9.90). In all cases they surpassed the national average (8.95). Similarly, the share of CIBER output published in journals well positioned according to their Journal Impact Factor was much higher than for the rest of the Spanish output. Almost half of the CIBER output was published in Q1 journals (47.32) while roughly above 20% of their output was published in D1 journals (22.41). CIBER 1 and 8 had the highest shares of Q1 papers (61.44 and 60.41) while CIBER 4 and again CIBER 1 showed the best performances regarding the D1 indicator (32.21 and 31.38).

**Table 3.** Output and impact indicators for each CIBER, the whole CIBER programme, Spain and Spanish Biomedical research excluding CIBER output for the 2005-2011 time period.

| | P | C | C/P | %Q1 | %D1 |
|---|---|---|---|---|---|
| **CIBER 1** | 4411 | 43666 | 9.90 | 61.64 | 31.38 |
| **CIBER 2** | 4508 | 56035 | 12.43 | 57.54 | 30.30 |
| **CIBER 3** | 2880 | 31907 | 11.08 | 54.93 | 22.99 |
| **CIBER 4** | 4356 | 58176 | 13.36 | 57.67 | 32.21 |
| **CIBER 5** | 3630 | 45261 | 12.47 | 57.16 | 25.62 |
| **CIBER 6** | 3104 | 36822 | 11.86 | 54.06 | 28.67 |
| **CIBER 7** | 4171 | 49436 | 11.85 | 55.31 | 25.25 |
| **CIBER 8** | 1710 | 19044 | 11.14 | 60.41 | 29.47 |
| **CIBER 9** | 2284 | 21753 | 9.52 | 57.57 | 27.89 |
| **SPAIN** | 111583 | 998548 | 8.95 | 40.76 | 17.05 |
| **CIBER** | 28251 | 330131 | 11.69 | 47.32 | 22.41 |
| **non-CIBER** | 86452 | 707764 | 8.19 | 37.15 | 14.69 |

**Note:** There are CIBER publications which are not categorized in biomedical fields

CIBER output has had a relative publication growth of 6.1% for the whole period. It represents approximately above 25% of the national output in the biomedical fields. While its number of publications has increased annually during the study time period, it has not increased as much as the rest of the Spanish biomedical output. Figure 2 shows the annual growth of publications from CIBER researchers as well as the share they represent from the whole output. As observed, since 2006 there has been a recession on the number of publications with respect to the rest of the country.

**Figure 2.** Number of publications produced by members of the CIBER programme and share they represent from the total Spanish biomedical outcome during the 2005-2011 time period

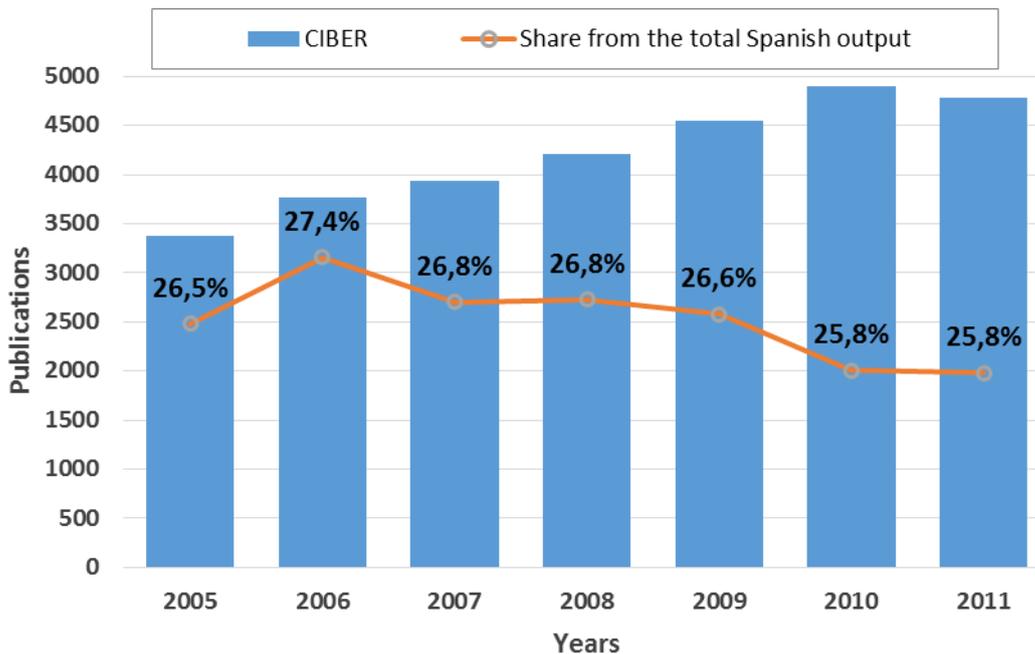

Table 4 shows number of CIBER papers published during the study time period in their most productive subject category. As observed, in regard with the national output, they represent below 50% of the total production. However, the citation impact and visibility of the CIBER output is much higher than the overall. The only exception can be found on the average number of citation received per paper for CIBER 7 (11.87 for CIBER papers and 12.83 for the Spanish output).

**Table 4.** Output and impact indicators for each of the JCR categories with the highest output per CIBER centre compared with the overall Spanish output during the 2005-2011 time period.

| CIBER centre | JCR Main output category | P | | %Q1 | | %D1 | | C/P | |
|---|---|---|---|---|---|---|---|---|---|
| | | CIBER | %SPAIN | CIBER | SPAIN | CIBER | SPAIN | CIBER | SPAIN |
| #1 | **Biomedical Engineering** | 467 | 34.09 | 47.97 | 40.07 | 24.20 | 16.57 | 8.04 | 6.57 |
| #2 | **Public, Environmental...Health** | 913 | 32.44 | 55.64 | 36.00 | 23.88 | 13.50 | 8.49 | 6.13 |
| #3 | **Endocrinology & Metabolism** | 690 | 20.25 | 47.10 | 40.82 | 14.49 | 12.59 | 13.52 | 11.20 |
| #4 | **Gastroenterology & Hepatology** | 1588 | 43.18 | 58.19 | 38.91 | 38.92 | 24.36 | 14.67 | 9.63 |
| #5 | **Neurosciences** | 1452 | 19.78 | 43.53 | 33.72 | 16.39 | 12.91 | 12.90 | 9.80 |
| #6 | **Respiratory System** | 704 | 29.72 | 50.14 | 38.08 | 40.20 | 24.19 | 13.15 | 9.34 |
| #7 | **Genetics & Heredity** | 846 | 17.88 | 46.57 | 40.37 | 19.74 | 16.59 | 11.87 | 12.83 |
| #8 | **Endocrinology & Metabolism** | 468 | 13.73 | 53.63 | 40.82 | 30.13 | 12.59 | 12.68 | 11.20 |
| #9 | **Psychiatry** | 1292 | 40.92 | 56.27 | 46.40 | 26.70 | 17.68 | 10.03 | 8.53 |

*3.2. Internal collaboration patterns in the CIBER programme*

In this section we focus on the collaboration trends developed within research groups in the CIBER programme during the study time period. For this, table 3 and figure 4 compare two time periods of two years each. The first time period has to do with the year prior to the establishment of the CIBER centre and its first year while the second period focuses on the 2010-2011 period. If we focus on the network analysis indicators (Co and D) we observe and increase for all CIBER centres. The greatest increase considering the share of groups forming part of the main component (Co) is for CIBER 1, where the co-authorship network was loosely tied at the beginning of the time period with just 31.9% of the research groups connected through the main component and has ended in the last two years with 68.1% of the nodes included in the main component. Regarding collaboration between researchers affiliated to different institutional classes, the evolution is not as significant and shares are similar or even lesser (as in CIBER 8 which descends from 6.8% to 3.1%).

Despite the increase in collaboration between CIBER research groups during the study time period, the share of papers done in collaboration is significantly low (under 20% of publications for all CIBER centres). Figure 3 shows the share of paper authored by one or more CIBER research groups. As observed, CIBER 9 is the centre with a higher share of collaborative papers while research groups in CIBER 1 co-authored papers with each other for less than 5% of their total output.

**Table 5.** Internal collaboration indicators by CIBER centre for two time periods: 2005-2006 and 2010-2011. For CIBER 8 and 9 the initial period is 2006-2007.

| | **Periods** | **Co** | **D** | **IC** |
|---|---|---|---|---|
| **CIBER 1** | 2005-2006 | 31,9 | 0,05 | 5,80 |
| | 2010-2011 | 68,1 | 0,1 | 8,22 |
| **CIBER 2** | 2005-2006 | 76,6 | 0,16 | 17,65 |
|  | 2010-2011 | 91,5 | 0,24 | 22,28 |
|  | 2005-2006 | 66,7 | 0,25 | 20,42 |
| **CIBER 3** | 2010-2011 | 88,9 | 0,49 | 20,48 |
|  | 2005-2006 | 76,7 | 0,12 | 7,17 |
| **CIBER 4** | 2010-2011 | 88,3 | 0,19 | 7,98 |
|  | 2005-2006 | 78,1 | 0,23 | 11,30 |
| **CIBER 5** | 2010-2011 | 93,8 | 0,27 | 16,58 |
|  | 2005-2006 | 78,1 | 0,23 | 4,16 |
| **CIBER 6** | 2010-2011 | 93,8 | 0,27 | 5,19 |
|  | 2005-2006 | 59,4 | 0,11 | 7,19 |
| **CIBER 7** | 2010-2011 | 84,8 | 0,17 | 7,08 |
| **CIBER 8** | 2006-2007 | 65,5 | 0,16 | 6,79 |
|  | 2010-2011 | 79,3 | 0,19 | 3,10 |
| **CIBER 9** | 2006-2007 | 88,5 | 0,36 | 8,67 |
|  | 2010-2011 | 100,0 | 0,61 | 9,43 |

In figure 4 we take another perspective and analyse collaboration between CIBER research groups and institutional classes according to the share of papers they produce in Q1 journals by CIBER centre (IC). In general terms, a larger share of collaborative papers are published in top journals, while the share of non-collaborative papers published in Q1 journals is lower. The only exception can be found for CIBER 4 where the share of IC papers was lower in both periods analysed (2005-2006: 43.8% IC vs. 47.5% No IC; 2010-2011: 49.3% IC vs. 53.3% No IC).

**Figure 3.** Share of research output by CIBER centre according to the number of research groups collaborating

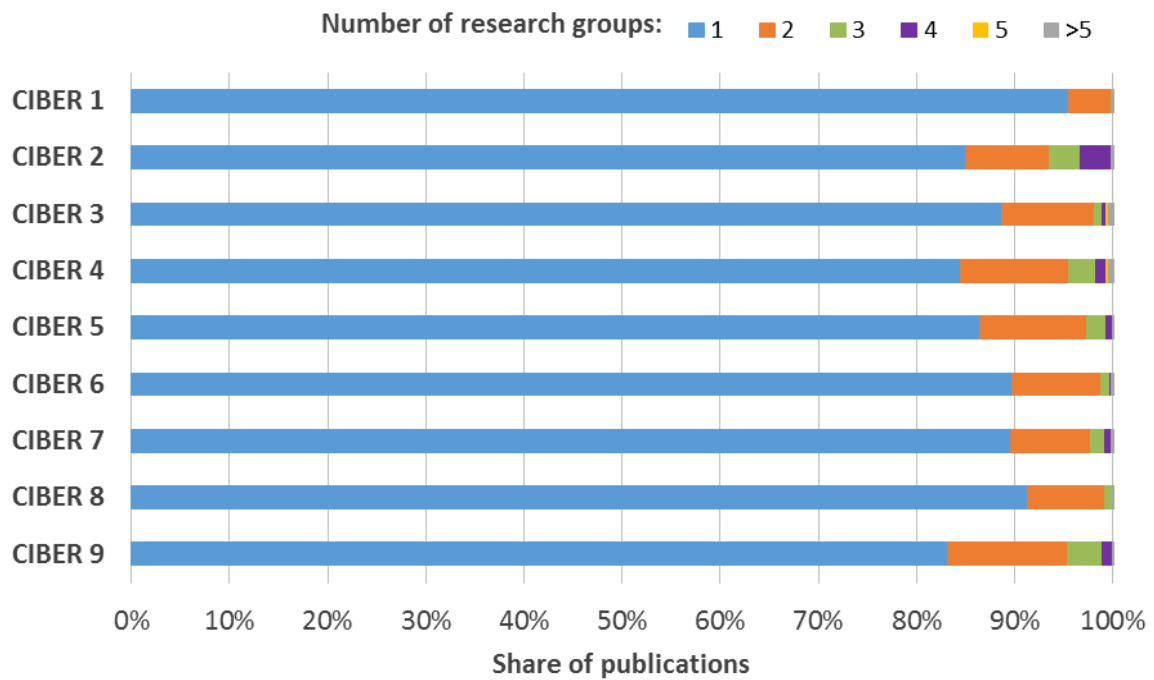

We must note that the share of No IC and No Collaboration Q1 papers was lower than 60% for all CIBER centres in both time periods. On the other hand, more than 60% of IC papers were published in Q1 journals for two centres (CIBER 6 and 9) in the first period, and seven centres (CIBER 1, 2, 3, 5, 7, 8 and 9) in the second period. With regard to papers authored by more than one CIBER group, the share of Q1 papers was again higher for all CIBER centres except for CIBER 4 (2005-2006: 42.3% IC vs. 49.3% No IC; 2010-2011: 50.0% IC vs. 50.8% No IC). The share of papers authored by research groups which did not collaborate with any other CIBER research group did not reach 60% Q1. In the first period, CIBER 5 reaches 60.2% of Q1 collaborative papers, while in the second period five centres surpass such threshold (CIBER 2, 5, 7, 8 and 9).

**Figure 4.** Share of Q1 publications by type of collaboration: different institutional categories vs. single institutional category, collaborative CIBER publications vs. non-collaborative CIBER publications at the beginning and end of the study time period. Beginning time period 2005-2006. End time period 2010-2011. * Beginning time period for CIBER centres 8 and 9 is 2006-2007.

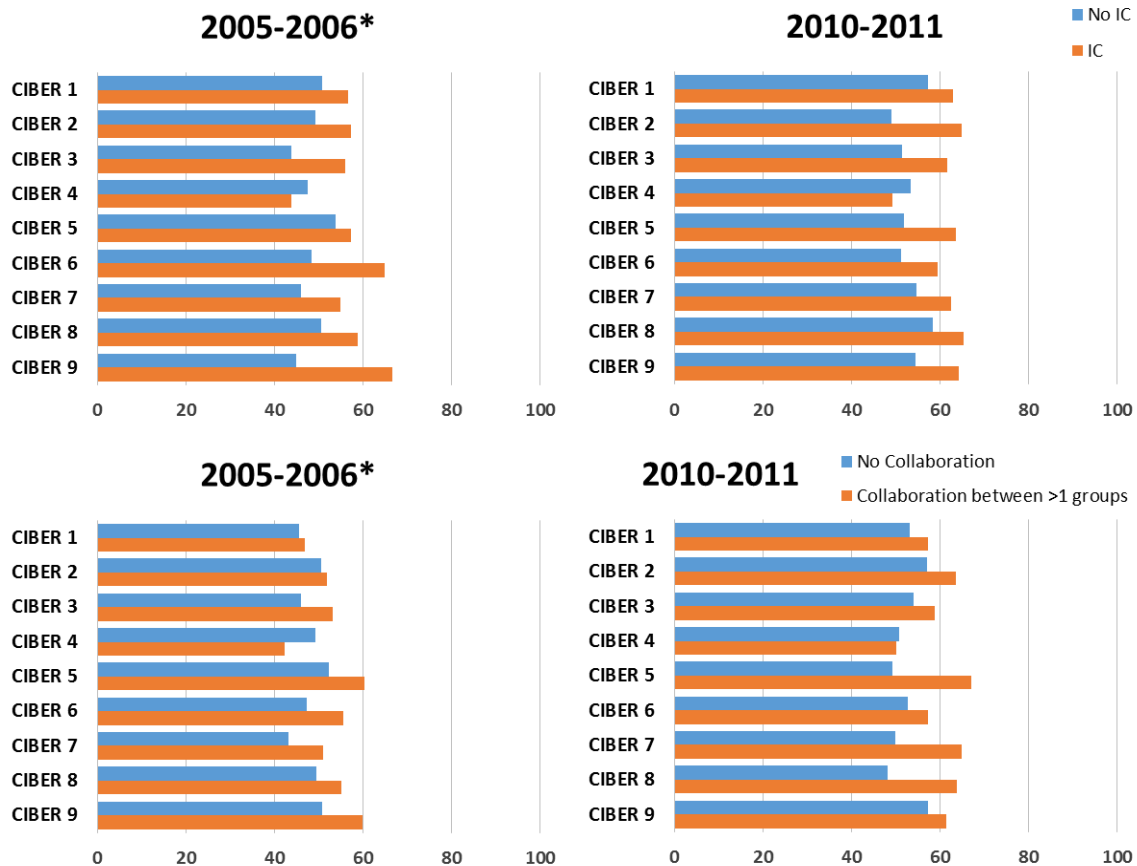

*3.3. Funding acknowledgment and impact*

In this section we show results regarding the share of papers where researchers include their CIBER information in the address field. 54.5% of the CIBER output could be identified as such through the address information; however, there are significant differences between centres (Table 6). While 75.2% of the papers produced by CIBER 9 were signed as such, 38.5% of the total output of CIBER 8 included such information. Indeed, for most centres (CIBER 1, 3, 4, 5, 6 and 8), less than half of their production were signed as CIBER.

If we focus only on papers published in Q1 journals, such share increases significantly to values between 56.9% (CIBER 3) and 66.7% of signed papers (CIBER 8). However such increase is not perceived for D1 publications, where the share of signed papers drops for all CIBER centres to shares between 20 and 30%. However, we do observe an increasing trend of the number of papers signed as CIBER since the launch of the programme (Figure 5). In this regard, we observe that there is an increasing awareness on behalf of researchers to acknowledge their membership to this research policy initiative. It is not until four years after the launch of the programme that half of the papers produced by CIBER researchers start to be signed as such. Here we see again a relation between the journal's position according to its Impact Factor and the share of papers signed as CIBER. Although the share of papers signed published in D1 journals is relatively low (Table 6), if we look at the opposite the trend is quite similar to that observed with regard to Q1 publications, that is, an important share (nearly 60% for 2011) of D1 publications indicate in their address that they are affiliated to a CIBER centre.

**Table 6.** Number of papers signed in the address field as CIBER for all publications 1st quartile publications and 1st decile publications by CIBER centre. Time period 2005-2011

|         | P     | $P_{signed}$ | $Q1_{signed}$ (%) | $D1_{signed}$ (%) |
|---------|-------|--------------|-------------------|-------------------|
| **CIBER 1** | 4411  | 1850         | 1131 (61.1)       | 516 (27.9)        |
| **CIBER 2** | 4508  | 2312         | 1337 (57.8)       | 588 (25.4)        |
| **CIBER 3** | 2880  | 1427         | 806 (56.9)        | 272 (19.1)        |
| **CIBER 4** | 4356  | 2107         | 1370 (65.0)       | 647 (30.7)        |
| **CIBER 5** | 3630  | 1668         | 1008 (60.4)       | 399 (23.9)        |
| **CIBER 6** | 3104  | 1394         | 824 (59.1)        | 396 (28.4)        |
| **CIBER 7** | 4171  | 2248         | 1296 (57.7)       | 523 (23.3)        |
| **CIBER 8** | 1710  | 658          | 439 (66.7)        | 174 (26.4)        |
| **CIBER 9** | 2284  | 1718         | 1087 (63.3)       | 481 (28.0)        |
| **TOTAL**   | 28251 | 15382        | 9298 (54.5)       | 3996 (26.0)       |

**Figure 5.** Share of papers signed as CIBER in the address field by year from the total CIBER output, from 1st quartile publications and from 1st decile publications. Time period 2005-2011

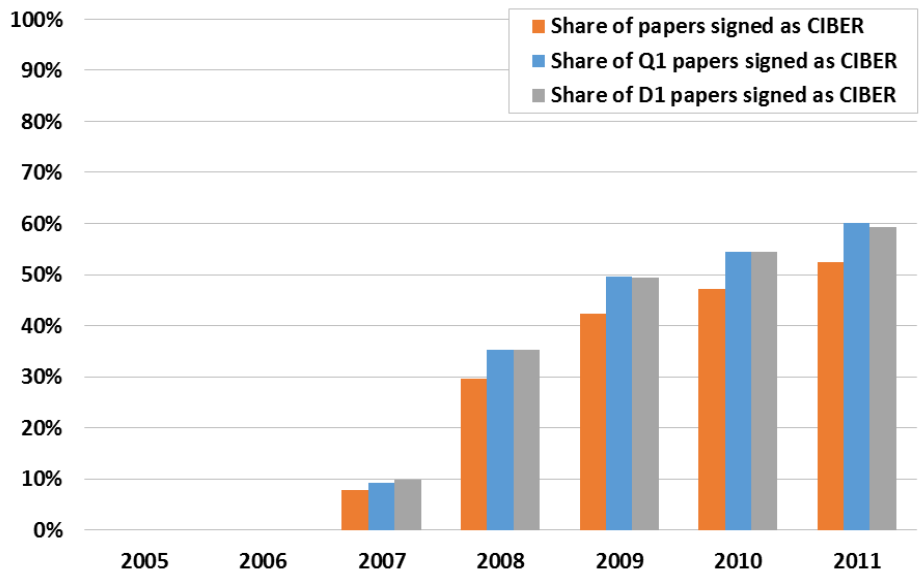

## 4. Discussion and concluding remarks

This paper analyses a research programme which intends to enhance translational research in the biomedical sciences in Spain. Among other goals, it aims at promoting collaboration between Spanish research teams. As observed in subsection 3.1, CIBER output represents an important share of Spanish output (25.3%), playing a key role on the visibility of Spanish research. In terms of citation impact and visibility the findings show that CIBER output is clearly higher than non-CIBER output in the biomedical fields. However, one may question if this is due to a good research groups selection or as a result of the interactions produced between its members.

To analyse this, subsection 3.2 focuses on the collaboration patterns of CIBER groups. We analyse internal collaboration at two levels: between research groups affiliated to the programme and between researchers belonging to different institutional classes; as goal of the programme is to enhance translational research through collaboration. In this regard, we observe that collaboration between research groups is generally low; however, the network is much more connected than it was before the programme launched. In this regard, one may hypothesize that the CIBER initiative has allowed research groups to connect with each other but has not succeeded on strengthening such connections. In this regard, maybe a wider study period may be necessary to confirm such hypothesis.

When comparing the effect of collaboration on impact we observe that, as confirmed in other studies (Bordons, Aparicio, & Costas, 2013), collaboration is an important variable that leads to higher visibility (figure 4). However, this should not be a capital issue for this programme as it is acknowledged elsewhere that clinical research may have a lower citation impact than basic research (van Eck, Waltman, van Raan, Klautz, & Peul, 2013). Although out of the scope of this study, here we must emphasize the need to analyse if any other types of collaboration have emerged in hospitals such as organising workshops, the publications of clinical guidelines, etc.

From a bibliometric perspective, an interesting aspect analysed is whether researchers acknowledge or not their affiliation to a CIBER centre. This aspect has already been discussed elsewhere (Morillo et al., 2015), however never for the whole programme and using a *bottom up* approach with external validation. This allows us to have an accurate picture of the behaviour researchers have with regard to their funding bodies. In this regard, we observe a similar share of papers that include the CIBER information in authors' affiliation to that reported by Morillo et al. (2015) for CIBER centres 4 and 9 (Hepatic and Digestive Diseases and Mental Health). In their study, they analyse the 2008-2011 time period with a greater increase on the share of signed papers (Figure 5).

The reasons way researchers choose to sign or not a paper as CIBER members may be due to how they value the programme as suggested by the heads of the centres (Cabezas-Clavijo et al., 2015). However, there is clearly a period in which such acknowledgment can take place. In this sense, maybe by analysing the share of publications signed as CIBER could be a good strategy to determine the time period needed for the programme to start to give results. In this regard, a follow up study would be desirable in order to analyse if the increasing share of signed papers stabilises over time and how many years since the launch of the programme are needed for this to happen.

**References**

Barjak, F., Es-Sadki, N. & Arundel, A. (2015). 'The effectiveness of policies for formal knowledge transfer from European universities and public research institutes to firms'. *Research Evaluation*, 24/1, 4–18.

Beaver, D. & Rosen, R. (1978). 'Studies in scientific collaboration'. *Scientometrics*, 1/1, 65–84.

Bordons, M., Aparicio, J. & Costas, R. (2013). 'Heterogeneity of collaboration and its relationship with research impact in a biomedical field'. *Scientometrics*, 96/2, 443–466.

Butler, D. (2008). 'Translational research: Crossing the valley of death'. *Nature*, 453/7197, 840–842.

Cabezas-Clavijo, Á., Robinson-Garcia, N. & Jimenez-Contreras, E. (2015). *¿Ha mejorado el programa CIBER la investigación biomédica española?: Una valoración a partir de las opiniones de sus responsables*.

Caron, E., & van Eck, N. J. (2014). 'Large scale author name disambiguation using rule-based scoring and clustering'. In *19th International Conference on Science and Technology Indicators. 'Context counts: Pathways to master big data and little data'* (pp. 79–86). CWTS-Leiden University Leiden.

Costas, R., & van Leeuwen, T. N. (2012). 'Approaching the 'reward triangle': General analysis of the presence of funding acknowledgments and 'peer interactive communication' in scientific publications'. *Journal of the American Society for Information Science and Technology*, 63/8, 1647–1661.

Delgado-Rodríguez, M. (2012). 'CIBERESP: un instrumento para la promoción y la difusión de la investigación de excelencia española'. *Gaceta Sanitaria*, 26/5, 393–394.

Demotes-Mainard, J., & Ohmann, C. (2005). 'European Clinical Research Infrastructures Network: promoting harmonisation and quality in European clinical research'. *The Lancet*, 365/9454, 107–108.

De Pablo, F., & Arenas, J. (2008). 'Introducción al Plan Nacional de Investigación, Desarrollo e Innovación 2008-2011: la acción estratégica en salud'. *Medicina Clínica*, 130/6, 223–227.

Glänzel, W. (2001). 'National characteristics in international scientific co-authorship relations'. *Scientometrics*, 51/1, 69–115.

Katz, J. S. & Martin, B. R. (1997). 'What is research collaboration?' *Research Policy*, 26/1, 1–18.

Lander, B. & Atkinson-Grosjean, J. (2011). 'Translational science and the hidden research system in universities and academic hospitals: A case study'. *Social Science & Medicine*, 72/4, 537–544.

Lee, S. & Bozeman, B. (2005). 'The impact of research collaboration on scientific productivity'. *Social Studies of Science*, 35/5, 673–702.

Luwel, M. & van Wijk, E. (2014). 'The 'translational' effect in medical journals: Bridging the gap?' *Research Evaluation*, 19/1, 19-29.

Méndez-Vásquez, R. I., Suñén-Pinyol, E., Olivé-Vázquez, G., Cervelló-Gonzáles, R., & Camí, J. (2009). *Caracterización bibliométrica de la producción en Epidemiología y Salud Pública (EPI-SP) de España, 1997–2006*. Barcelona. Retrieved from http://bac.fundaciorecerca.cat/ciberesp/docs/Bibliometria_ CIBERESP.pdf

Molas-Gallart, J., D'Este, P., Llopis, Ó. & Rafols, I. (2014). *Towards an alternative framework for the evaluation of translational research initiatives*. INGENIO (CSIC-UPV). Retrieved from http://www.ingenio.upv.es/sites/default/files/working-paper/2014-03.pdf

Morillo, F., Costas, R. & Bordons, M. (2015). 'How is credit given to networking centres in their publications? A case study of the Spanish CIBER research structures'. *Scientometrics*, 103/3, 923-938.

Morillo, F., Díaz-Faes, A. A., González-Albo, B. & Moreno, L. (2014). 'Do networking centres perform better? An exploratory analysis in Psychiatry and Gastroenterology/Hepatology in Spain'. *Scientometrics*, 98/2, 1401–1416.

Rettig, R. A., Schechter, A. N., & Perlman, R. L. (2004). 'Editors' Introduction: Why is Revitalizing Clinical Research So Important, Yet So Difficult?' *Perspectives in Biology and Medicine*, 47/4, 476–486.

Rey-Rocha, J., & Martín-Sempere, M. J. (2012). 'Generating favourable contexts for translational research through the incorporation of basic researchers into hospitals: The FIS/Miguel Servet Research Contract Programme'. *Science and Public Policy*, 39/6, 787–801.

van Eck, N. J., Waltman, L., van Raan, A. F. J., Klautz, R. J. M. & Peul, W. C. (2013). 'Citation Analysis May Severely Underestimate the Impact of Clinical Research as Compared to Basic Research'. *PLoS ONE*, 8/4, e62395.

van Leeuwen, T. N. (2007). 'Modelling of bibliometric approaches and importance of output verification in research performance assessment'. *Research Evaluation*, 16/2, 93–105.